\documentclass[10pt,aps,prl,twocolumn,superscriptaddress,preprintnumbers,floatfix,notitlepage]{revtex4-1}
\usepackage{hyperref}
\hypersetup{
    colorlinks=true,
    linkcolor=blue,
    filecolor=magenta,      
    urlcolor=blue,
}
\usepackage{amsmath,amsfonts,amssymb}
\usepackage{fancyhdr}
\usepackage{graphicx}
\usepackage{xspace}
\usepackage{rotating}
\usepackage[normalem]{ulem}
\usepackage{braket}
\usepackage{verbatim}
\usepackage{xcolor}
\usepackage{hyperref}
\usepackage[utf8]{inputenc}

\def\lsim{\mathrel{\raise.3ex\hbox{$<$\kern-.75em\lower1ex\hbox{$\sim$}}}}
\def\gsim{\mathrel{\raise.3ex\hbox{$>$\kern-.75em\lower1ex\hbox{$\sim$}}}}

\newcommand{\be}{\begin{equation}}
\newcommand{\ee}{\end{equation}}
\newcommand{\bea}{\begin{equation}\begin{aligned}}
\newcommand{\eea}{\end{aligned}\end{equation}}

\begin{document}
\begin{flushright}
KCL-PH-TH/2020-53, CERN-TH-2020-150
\end{flushright}

\title{Cosmic String Interpretation of NANOGrav Pulsar Timing Data}

\author{John~Ellis,}
\email[]{john.ellis@cern.ch}
\affiliation{Kings College London, Strand, London, WC2R 2LS, United Kingdom}
\affiliation{Theoretical Physics Department, CERN, Geneva, Switzerland}
\affiliation{National Institute of Chemical Physics \& Biophysics, R\"avala 10, 10143 Tallinn, Estonia}

\author{Marek Lewicki}
\email[]{marek.lewicki@kcl.ac.uk}
\affiliation{Kings College London, Strand, London, WC2R 2LS, United Kingdom}
\affiliation{Faculty of Physics, University of Warsaw ul.\ Pasteura 5, 02-093 Warsaw, Poland}

\begin{abstract}Pulsar timing data used to provide upper limits on a
possible stochastic gravitational wave background (SGWB). However, the NANOGrav Collaboration has recently reported strong evidence for a stochastic common-spectrum process, which we interpret as a
SGWB in the framework of cosmic strings. The possible NANOGrav
signal would correspond to a string tension $G\mu \in (4 \times 10^{-11},  10^{-10}) $ at the 68\% confidence level, with a different frequency dependence from supermassive black hole mergers.
The SGWB produced by cosmic strings with such values of $G\mu$ would be beyond the reach of LIGO, but could be measured by other planned and proposed detectors such as SKA, LISA, TianQin, AION-1km, AEDGE, Einstein Telescope and Cosmic Explorer.
\end{abstract}

\maketitle

%%%%%%%%%%%%%%%%%%%%%%%%%%%%%%%%%%%%%%%%%%%%%%%%%%%%%%%%
{\bf Introduction:} Stimulated by the direct discovery of gravitational waves (GWs) by the LIGO and Virgo Collaborations~\cite{Abbott:2016blz,Abbott:2016nmj,Abbott:2017vtc,Abbott:2017gyy,Abbott:2017oio,TheLIGOScientific:2017qsa,Abbott:2020khf,Abbott:2020tfl} of black holes and neutron stars
at frequencies $f \gtrsim 10$~Hz, there is widespread interest in experiments exploring other parts
of the GW spectrum. Foremost among these are pulsar timing array (PTA) experiments, which are
sensitive to GWs with frequencies $f \lesssim 1$/yr. PTA experiments probe the possible existence of a
stochastic GW background (SGWB), as might be generated by very different physical phenomena such as astrophysical sources of GWs, 
e.g., the mergers of supermassive black hole (SMBHs), or cosmological sources, e.g., cosmic strings.

Aggregating pulsar measurements for over a decade, the EPTA~\cite{Lentati:2015qwp}, PPTA~\cite{Shannon:2015ect} and NANOGrav~\cite{Arzoumanian:2018saf} PTA experiments have pushed
their sensitivities down to an energy density $\Omega_{\rm GW} h^2 \lesssim 10^{-9}$ over frequencies in
the range $f \in (2.5 \times 10^{-9},1.2\times  10^{-8})$~Hz. Until recently, there has been no indication of a
positive signal above background. However, a recent NANOGrav analysis of 12.5~yrs of pulsar timing
data~\cite{Arzoumanian:2020vkk} reports strong evidence for a stochastic common-spectrum process  that may be interpreted as a GW signal with amplitude
$A \sim {\cal O}(10^{-15})$ at $f \sim 1$/yr.
The NANOGrav Collaboration notes that 
this signal is in apparent tension with previous upper limits on the SGWB in this frequency range,
but argues that this is not real, but reflects its improved treatment of the intrinsic pulsar red noise.
The NANOGrav signal does not exhibit significant monopole or dipole correlations,
as might arise, e.g., from reference clock or solar-system ephemeris systematics, respectively.
On the other hand, neither does the signal exhibit significant quadrupole correlations, which would have been a
``smoking gun" for a GW background, and the NANOGrav Collaboration does not claim a detection of GWs.

Nevertheless, we are emboldened to explore the implications of this possible SGWB detection by
NANOGrav for cosmic string models, discussing how experiments could confirm or disprove
such an interpretation. Upper limits on the SGWB are often quoted assuming a spectrum described by a GW abundance proportional to $f^{2/3}$, 
as expected for SMBH mergers~\cite{Phinney:2001di}.
However, the cosmic string GW spectrum is not a simple power law, but is convex with an amplitude and a frequency-dependent slope that depend on the parameter, $G \mu$, where $G$ is the Newton constant of gravitation and $\mu$ is the string tension.
Any limit (or estimate) of $G \mu$ from any specific experiment must take into account take into account the appropriate
slope parameter, which is in general $\ne 2/3$ in the characteristic frequency measurement range. Once an
allowed (interesting)  value of $G \mu$ has been identified, however, the cosmic string prediction for the
magnitude and spectral shape of the SGWB is then fixed as a function of frequency, and can then be compared
with the sensitivities of other experiments.

In this paper we calculate the effective slope parameter {for the timing-residual cross-power spectral density $\gamma$ (which translates to $\gamma=5-\beta$ for $\Omega\propto f^\beta$) for frequencies in the range $(2.5 \times 10^{-9}, 1.2 \times 10^{-8})$~Hz
used  in~\cite{Arzoumanian:2020vkk} to make a single-power fit to the NANOGrav 12.5~yr data.}
The best fit to the NANOGrav data is shown 
as an orange dashed line in the left panel of Fig.~1 of~\cite{Arzoumanian:2020vkk}, and the 68\% and 95\%
CL ranges in the $(\gamma, A)$ plane are shown as orange dashed and dotted ellipses in the right panel of 
Fig.~1 of~\cite{Arzoumanian:2020vkk}. We find that the cosmic string model gives a better fit than does a single
power law with $\gamma = 13/3$ as suggested by models of SMBH mergers: the one-parameter cosmic string 
prediction crosses the 68\% CL ellipse, whereas the $\gamma = 13/3$ line passes outside it
though within the 95\% ellipse. The GW spectra
predicted by the cosmic string model for $G \mu \in (2 \times 10^{-11}, 2 \times 10^{-10})$, the range where it
lies within the NANOGrav 12.5~yr 95\% CL region in the $(\gamma, A)$ plane, are all completely compatible 
with the EPTA upper limit,
{although some tension with with the PPTA results remains in the upper part of our range.}
The cosmic
string predictions are well within the estimated reaches of the SKA~\cite{Janssen:2014dka}, LISA~\cite{Bartolo:2016ami,Caprini:2019pxz}, TianQin~\cite{Luo:2015ght,Mei:2020lrl},
AEDGE~\cite{Bertoldi:2019tck}, AION-1km~\cite{Badurina:2019hst}, 
ET~\cite{Punturo:2010zz,Hild:2010id} and CE~\cite{Evans:2016mbw} experiments, 
but beyond the present and estimated future sensitivities of the
LIGO~\cite{TheLIGOScientific:2014jea,Thrane:2013oya,TheLIGOScientific:2016wyq,LIGOScientific:2019vic} experiment.

{\bf GW spectrum from cosmic strings:} Cosmic strings are one-dimensional stable objects described by their characteristic tension $\mu$. They are a common prediction of many extensions of the Standard Model~\cite{Jeannerot:2003qv,King:2020hyd} featuring a $U(1)$ symmetry-breaking phase transition in the early universe~\cite{Nielsen:1973cs}. They can also arise in superstring theory as cosmologically-stretched fundamental strings~\cite{Dvali:2003zj,Copeland:2003bj}. 
We focus mostly on the former case, for which the inter-commutation probability $p$ (the probability that strings reconnect in a different way after crossing) takes the value $p=1$, and comment on this choice towards the end of the following Section.

We use a simple method of computation of the GW spectrum from a cosmic string network following~\cite{Cui:2017ufi,Cui:2018rwi} {(for an overview, see~\cite{Auclair:2019wcv})}. We utilise the Velocity-dependent One-Scale
(VOS) model~\cite{Martins:1995tg, Martins:1996jp,Martins:2000cs},
assuming that the length of a loop produced by the network $\ell$ at time  $t_i$ evolves as
\be
\ell = \alpha_\ell t_i - \Gamma G\mu(t-t_i) \ ,
\label{eq:looplength}
\ee
where $G\mu$ is the string tension and $\alpha_\ell$ the initial loop size.
Following the guidance from recent numerical simulations~\cite{Blanco-Pillado:2013qja,Blanco-Pillado:2017oxo}, we focus on the largest loops produced by the network, fixing $\alpha_\ell=0.1$, as these dominate the GW emission. String loops emit at normal oscillation mode frequencies, allowing us to express the frequency measured today from mode $k$ with emission time $\tilde{t}$ as
\be
f = \frac{a(\tilde{t})}{a(t_0)}
\,
\frac{2k}{\alpha_\ell t_i - \Gamma G\mu(\tilde{t}-t_i)} \ ,
\label{eq:ftoday}
\ee
where $t_0$ is the current time.
The GW abundance can be computed as a sum over individual emission modes
\be
\label{2}
\Omega_{GW}^{CS}(f) =  \sum_{k=1}^\infty k \Gamma^{(k)} \Omega^{(k)}_{GW}(f)\, ,
\ee
where the total emission rate $\Gamma$ is found in simulations to have the value $\Gamma \simeq 50$~\cite{
Vilenkin:1981bx,Turok:1984cn,Quashnock:1990wv,Blanco-Pillado:2013qja,Blanco-Pillado:2017oxo}, and we assume that this is dominated by emission from cusps with $\Gamma^{(k)}  = \Gamma k^{-\frac{4}{3}}/(\sum_{m=1}^{\infty} m^{-\frac{4}{3}})$~\cite{Cui:2018rwi}. We truncate the sum in Eq.~(\ref{2}) at $10^3$ modes, beyond which we approximate it with an integral that guarantees good accuracy also for the high-frequency part of the spectrum~\cite{Cui:2019kkd,Blasi:2020wpy,Gouttenoire:2019kij}.
The contribution of each emission mode in Eq.~(\ref{2}) has the form (see~\cite{Cui:2018rwi} for details)
\begin{align}
\label{4}
\Omega^{(k)}_{GW}(f) &=\frac{16 \pi}{3H_0^2}\frac{(0.1)\, (G\mu)^2}{\alpha_\ell(\alpha_\ell+\Gamma G\mu)}\frac{1}{f} \\
\times & \int_{t_F}^{t_0}\!\!d\tilde{t}\,\;\frac{C_{eff}(t_i)}{t_i^{4}}
\left(\frac{a(\tilde{t})}{a(t_0)}\right)^5\left(\frac{a(t_i)}{a(\tilde{t})}\right)^3 \Theta(t_i-t_F) \ . \nonumber
\end{align}
 In evaluating the scale factor $a(t)$, we use the number of degrees of freedom predicted by the Standard Model as given by microMEGAS~\cite{Belanger:2018ccd}.
 The lower integration limit $t_F$ corresponds to the network formation time, which can be assumed to be an arbitrarily small number for our purposes, as it only controls the high frequency cut-off of the spectrum, whereas we are mostly interested in the low-frequency peak.~\footnote{In fact, in generic cases a much more important cut-off on the high-frequency end of the spectrum appears where particle emission becomes more important than GW emission~\cite{Auclair:2019jip,Matsunami:2019fss}.} We calculate the $C_{eff}$ factor controlling the loop number density in Eq.~(\ref{4}) using the velocity-dependent one-scale~(VOS)~\cite{Martins:1995tg,Martins:1996jp,Martins:2000cs,Avelino:2012qy,Sousa:2013aaa} model as in~\cite{Cui:2017ufi,Cui:2018rwi} which gives $C_{eff} = 5.4$ and $0.39$ during radiation and matter domination, respectively. These values agree quite well with the values predicted by recent numerical simulations~\cite{BlancoPillado:2011dq,Blanco-Pillado:2013qja,Blanco-Pillado:2017oxo,Blanco-Pillado:2019vcs,Blanco-Pillado:2019tbi}. Finally the additional factor $0.1$ comes from the same simulations, which find that only this fraction of energy goes into large loops that produce GWs efficiently, whereas the rest goes into the kinetic energy of small loops that is then lost to redshifting.

{\bf Connection with experimental results:}
The most recent experimental results from 12.5~yr of NANOGrav data~\cite{Arzoumanian:2020vkk} are expressed in terms of a generic power-law signal with characteristic strain given by
\be
h_{c}(f)=A\left(\frac{f}{f_{\mathrm{yr}}}\right)^{\alpha},
\ee
where $f_{\rm yr}=1{\rm yr}^{-1}$.
The abundance of gravitational waves has the standard form, which can also be recast as a power-law:
\be\label{eq:PLabundance}
\Omega(f)=\frac{2\pi^2}{3H_0^2}f^2 h_c(f)^2=\Omega_{yr}\left(\frac{f}{f_{\mathrm{yr}}}\right)^{\beta}=\Omega_{yr}\left(\frac{f}{f_{\mathrm{yr}}}\right)^{5-\gamma}, \,
\ee
where
\be
\Omega_{yr}=\frac{2\pi^2}{3H_0^2}A^2 f_{yr}^2\, .
\ee
The experimental analysis was cast in terms of the power law found in
the timing-residual cross-power spectral density $\gamma=3-2\alpha=5 -\beta$, and we adopt this notation.

\begin{figure}
\centering
\includegraphics[width=0.4\textwidth]{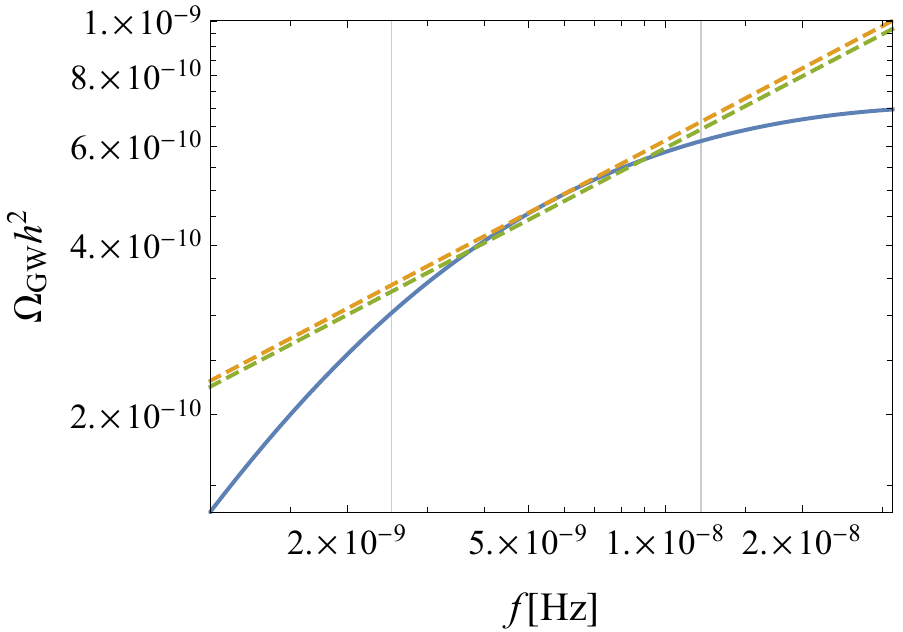}
\caption{\it Cosmic string spectra (solid blue curves) together with our fitted power laws for $G\mu=4\times 10^{-11}$.%, and $G\mu=10^{-10}$.
The green dashed lines show the results of numerically fitting the curves, while the orange lines result from the simple logarithmic derivative in Eq.~(\ref{eq:PLfit}). The thin grey lines indicate the frequency range of interest that was used in the NANOGrav linear fit. 
\label{fig:fits}}
\end{figure}

In order to make connection with the experimental results, we approximate the cosmic string spectra with power laws in the range of frequencies where the possible signal was observed. The simple power-law approximation used by NANOGrav~\cite{Arzoumanian:2020vkk} was fitted to 5 bins covering roughly $f \in (2.5\times 10^{-9},1.2\times 10^{-8})$~Hz, with the higher-frequency bins still seemingly dominated by noise in the data. To  estimate the prospective cosmic string signal for any given value of $G\mu$, we fit numerically a power law, see Eq.~(\ref{eq:PLabundance}), to the calculation of the spectrum described above in the range of interest. We show an example of this fit for $G\mu=4\times 10^{-11}$ in Fig.~\ref{fig:fits}. However, as also as we also see in the plot, we find that a very good approximation is obtained by simply taking a logarithmic derivative of our cosmic string spectrum to find the slope
\be\label{eq:PLfit}
\begin{split}
\gamma & =5-\left. \frac{d \log \Omega_{GW}^{CS}(f)}{d \log f}\right|_{f=f_*}, \\ A & =\sqrt{\frac{3H_0^2}{2\pi^2} \ \frac{\Omega_{GW}^{CS}(f_*)(f_{\rm yr}/f_*)^{5-\gamma}}{f_{\rm yr}^2}}
\end{split}
\ee
 at the reference frequency $f_*\approx 5.6 \times 10^{-9}$~Hz.

\begin{figure}
\centering
\includegraphics[width=0.499\textwidth]{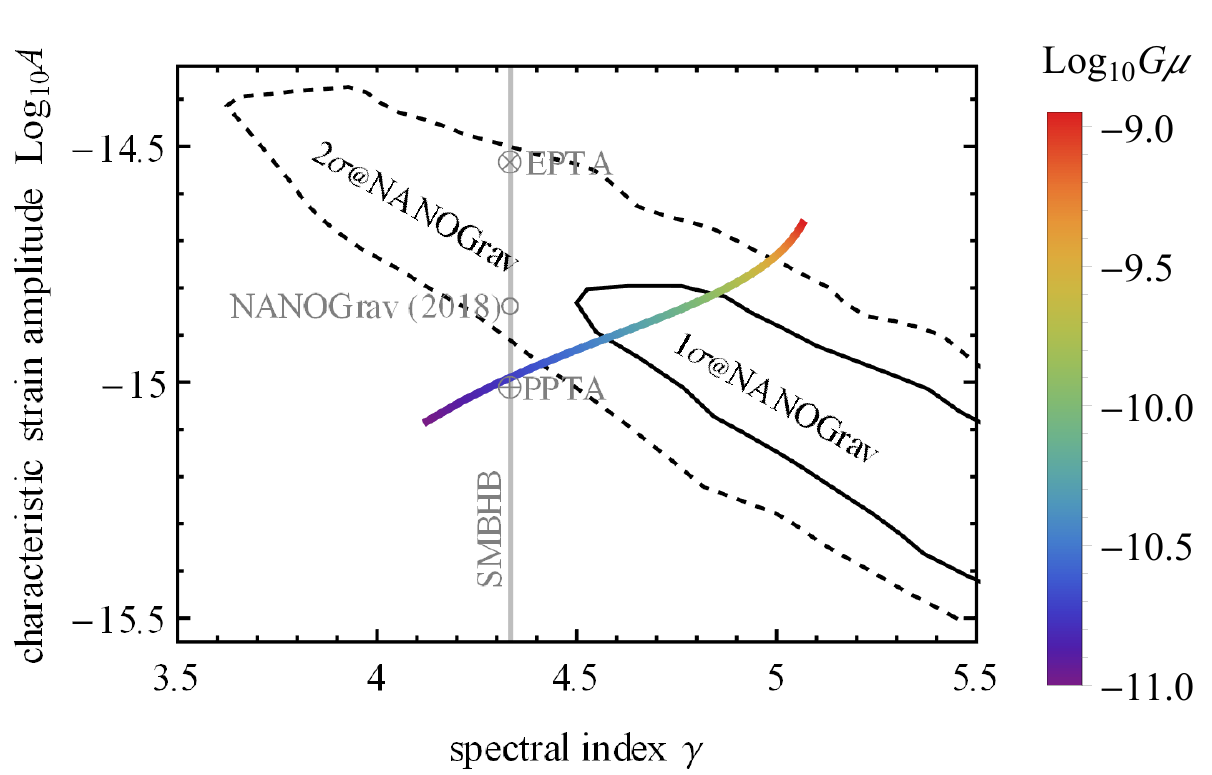}
\caption{\it The curve shows the slope $\gamma$ and amplitude $A$ of a power law signal approximating the calculated cosmic string spectra, see Eqs.~(\ref{eq:PLabundance}) and (\ref{eq:PLfit}), with $G\mu$ values indicated by the indicated rainbow colours in the indicated frequency range. The solid and dashed black lines indicate the $68\%$ and $95\%$ ranges of $(\gamma, A)$ fitted to their 12.5~yr data by the  NANOGrav collaboration~\cite{Arzoumanian:2020vkk}. The grey vertical line at $\gamma=13/3$ represents the slope expected for SMBH mergers~\cite{Phinney:2001di}, while the points on it mark the upper limits on the amplitude  from previously-reported pulsar timing data for that spectrum.
\label{fig:Agammaplot}}
\end{figure}

We show in Fig~\ref{fig:Agammaplot} the resulting values of $\gamma$ and $A$ for a range of $G\mu$ values of interest overlaid on the NANOGrav fit to their 12.5~yr data~\cite{Arzoumanian:2020vkk}. We find that values of the string tension $G\mu\in (4\times 10^{-11},10^{-10})$ give results within the 68\% CL range of the NANOgrav fit, while $G\mu\in (2\times 10^{-11},3\times 10^{-10})$ make predictions within the 95\% range. Interestingly, the cosmic string interpretation offers a slightly better fit than SMBH mergers, which predict $\gamma=13/3$ (shown as the vertical gray line in Fig.~\ref{fig:Agammaplot}) yielding a fit that is at best within the 95\% CL range but outside the 68\% range.

\begin{figure}
\centering
\includegraphics[width=0.49\textwidth]{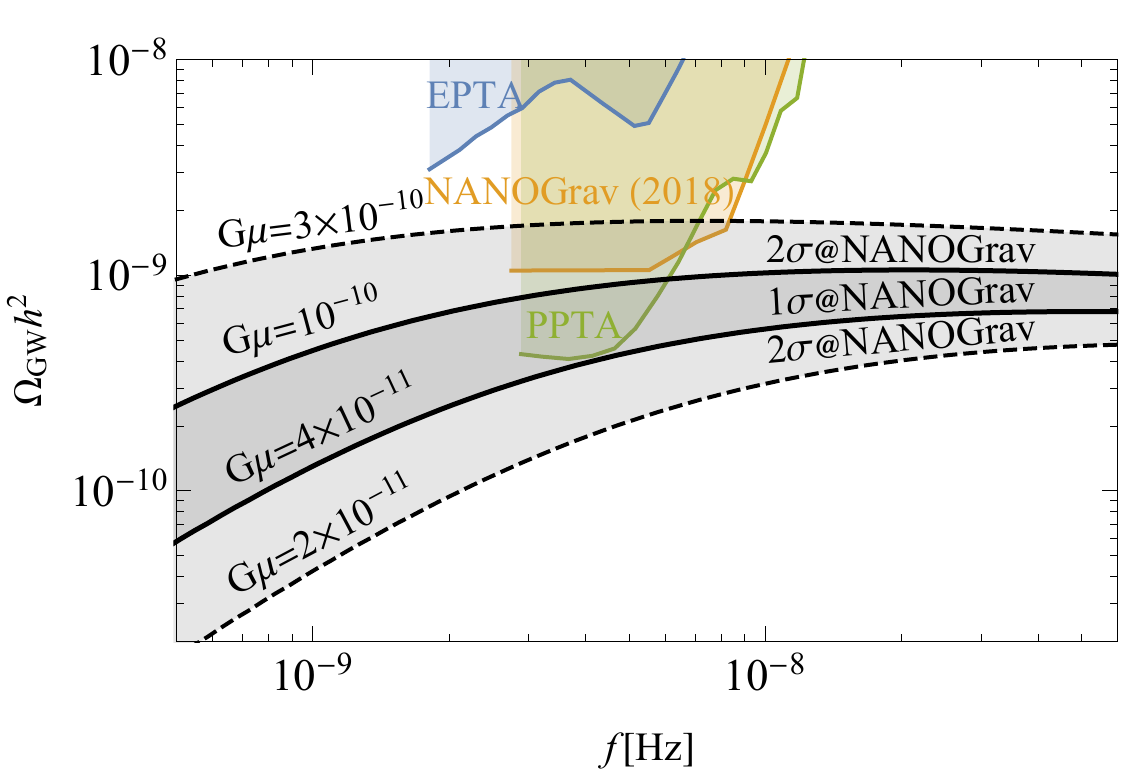}
\caption{\it Cosmic string spectra calculated for $f \in 5 \times 10^{-10}, 6 \times 10^{-8})$ with $G\mu\in (4\times 10^{-11},10^{-10})$ (between the solid black lines)  and $G\mu\in (2\times 10^{-11},3\times 10^{-10})$ (between the dashed black lines) that fit the NANOGrav 12.5~yr data within the 68\% and 95\% confidence levels, respectively. We also show previously reported bounds from PPTA~\cite{Shannon:2015ect}, EPTA~\cite{Lentati:2015qwp} and NANOGrav 11~yr data~\cite{Arzoumanian:2018saf}. 
\label{fig:OmegaPTAzoom}}
\end{figure}

The new NANOGrav 12.5~yr~\cite{Arzoumanian:2020vkk} results are in some tension with previous bounds from PPTA~\cite{Shannon:2015ect} and a previous NANOGrav analysis of their 11~yr data~\cite{Arzoumanian:2018saf}, though compatible with EPTA data~\cite{Lentati:2015qwp}. Fig.~\ref{fig:OmegaPTAzoom} shows a comparison of the older constraints with the cosmic string spectra that provide 68\% and 95\% CL fits to the NANOGrav 12.5~yr data. The apparent tension is also visible in Fig.~\ref{fig:Agammaplot}, which shows previous PPTA and NANOGrav upper limits on the amplitude of a $\gamma=13/3$ SMBH merger spectrum (vertical grey line) from the earlier pulsar timing data releases cited above. According to the NANOGrav collaboration~\cite{Arzoumanian:2020vkk}, their new analysis uses improved priors for the intrinsic pulsar red noise (see~\cite{Hazboun:2020kzd} for a recent discussion). Applying these new priors to older data would ease the previous constraints and tend to reduce the tension.

\begin{figure}
\centering
\includegraphics[width=0.49\textwidth]{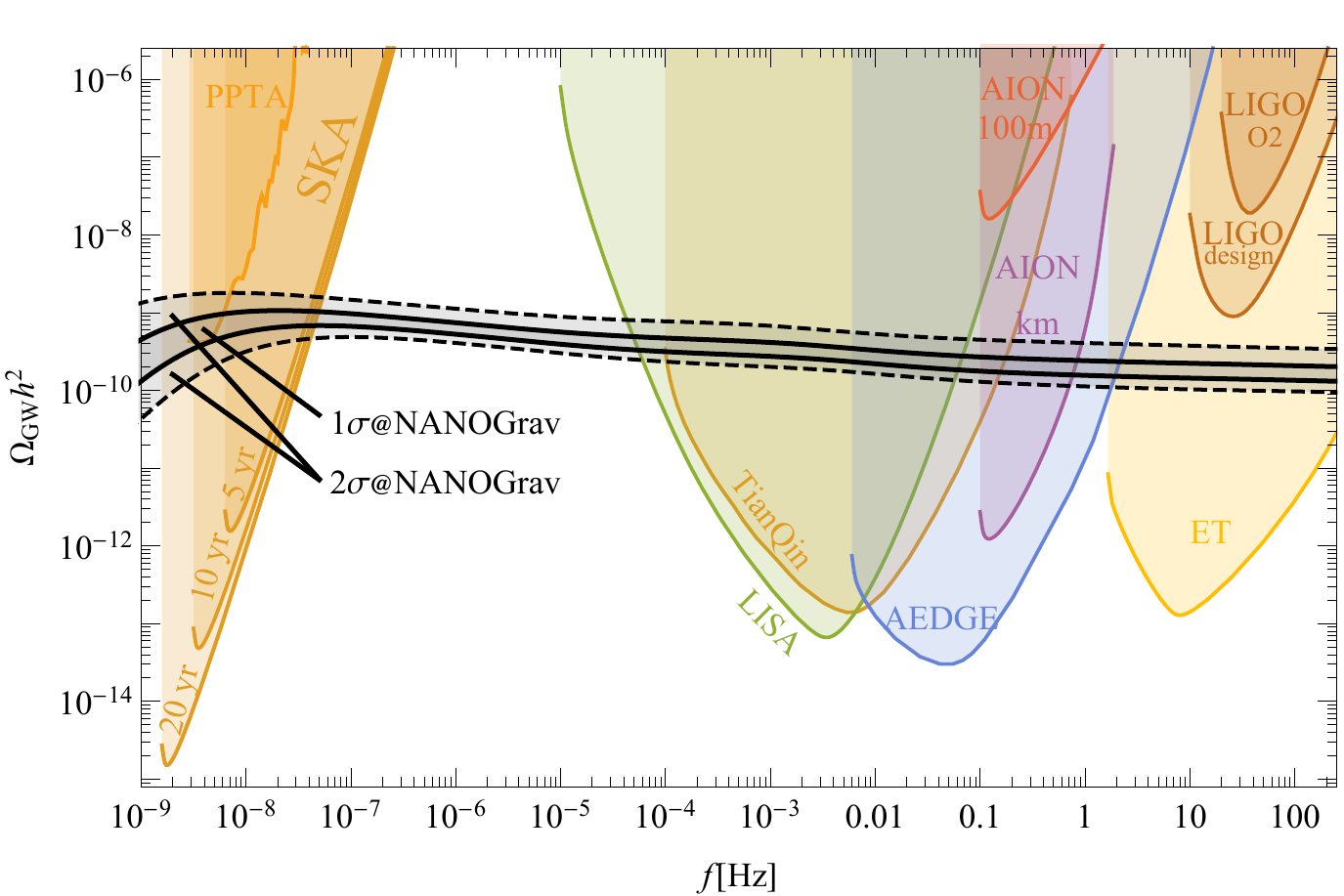}
\caption{\it Cosmic string spectra calculated for $f \in (10^{-9}, 200)$~Hz with $G\mu\in (4\times 10^{-11},10^{-10})$ (between the solid black lines)  and $G\mu\in (2\times 10^{-11},3\times 10^{-10})$ (between the dashed black lines) that fit the NANOGrav 12.5~yr data at the 68\% and 95\% confidence levels, respectively. We also show the current sensitivity of LIGO O2 as well as its design sensitivity, as well as the estimated reaches of the other planned and proposed experiments SKA, LISA, TianQin, AEDGE, AION, ET and CE. 
\label{fig:OmegaPTA}}
\end{figure}

Fig.~\ref{fig:OmegaPTA} shows the spectra that fit the new NANOGrav data  at the 68\% and 95\% CLs over an extended frequency range $f \in (10^{-9}, 200)$~Hz. We also show the current sensitivity of LIGO O2~\cite{LIGOScientific:2019vic} together with its design sensitivity goal~\cite{TheLIGOScientific:2014jea,Thrane:2013oya,TheLIGOScientific:2016wyq},
as well as the projected sensitivities of SKA~\cite{Janssen:2014dka} and the upcoming GW experiments LISA~\cite{Bartolo:2016ami,Caprini:2019pxz},  TianQin~\cite{Luo:2015ght,Mei:2020lrl}, AEDGE~\cite{Bertoldi:2019tck}, AION/MAGIS~\cite{Badurina:2019hst,Graham:2016plp,Graham:2017pmn} and ET~\cite{Punturo:2010zz,Hild:2010id}.
We see all the next-generation GW experiments should be able to observe cosmic string signals strong enough to fit the current NANOGrav data. However, LIGO would, unfortunately not be able to observe such a signal even after reaching its design sensitivity~\footnote{However, LIGO could potentially probe spectra fitting the  data in alternative models with additional features due, e.g., to modification of the spectrum by non-standard cosmological expansion~\cite{Cui:2017ufi,Cui:2018rwi,Gouttenoire:2019kij}, or cosmic string models featuring large production of very small scale loops~\cite{Lorenz:2010sm,Ringeval:2017eww,Auclair:2019zoz,Auclair:2020oww}.}.  

We have focused throughout this Section on cosmic strings that always interchange partners upon crossing, so that the inter-commutation probability $p=1$, though this could be reduced if the strings originated from superstring theory~\cite{Dvali:2003zj,Copeland:2003bj}. In a first approximation this just corresponds to the density of strings increasing as $p^{-1}$ for any given value of the tension, which leads to a similar increase in the amplitude of the GW signal~\cite{Sakellariadou:2004wq,Blanco-Pillado:2017rnf}. As a result, the cosmic string curve in Fig.~\ref{fig:Agammaplot} would simply move up in amplitude as $A\propto \sqrt{\Omega}\propto\sqrt{p^{-1}}$. Since the rainbow curve passes close to the top of the NANOGrav 68\% CL region, there is little scope for decreasing $p$ while maintaining consistency at the 68\% CL, with $\Omega_{\rm GW} h^2$ increasing by $< 50$\%. We note, however, that the dependence of the density on the probability can be milder than the simple $p^{-1}$ assumption~\cite{Avgoustidis:2005nv}, and that the final result is still a matter of debate~\cite{Auclair:2019wcv}, so that this conclusion may need to be relaxed.

Before proceeding to our conclusions, we first mention briefly other possible sources that could potentially fit the new NANOGrav data.
One possibility is SMBH mergers. However, their rate is uncertain and, as already noted, a simple model led to the prediction $\gamma = 13/3$~\cite{Phinney:2001di} that is apparently disfavoured by the NANOGrav data, though this is sensitive to the priors used in the data analysis~\cite{vanHaasteren:2008yh,Hazboun:2020kzd}. Another possibility is primordial inflation~\cite{Bartolo:2016ami,Caprini:2018mtu}, which leads generically to a flat spectrum with $\gamma=5$ whose magnitude is constrained by CMB measurements~\cite{Aghanim:2015xee} at $f_{CMB} \approx 10^{-17}$ Hz to be orders of magnitude below the amplitude of the observed signal. The inflationary spectrum would therefore require modification if it is to fit the observed abundance at PTA frequencies. This requires $\beta \simeq 0.68$~\cite{Lasky:2015lej}, which gives a spectrum at PTA frequencies with $\gamma \simeq 4.32$, a value very close to the SMBH merger prediction~\cite{Phinney:2001di} and again seemingly slightly disfavoured by the current data.
A third possibility is a signal from a first-order phase transition in the early universe. However, such a signal typically peaks at a much higher frequency~\cite{Caprini:2015zlo,Caprini:2019egz}. Lowering the frequency requires a transition at a lower temperature, which is possible only in a model with a hidden sector decoupled from the Standard Model~\cite{Breitbach:2018ddu}, since the frequency cannot be lowered by supercooling~\cite{Ellis:2018mja,Ellis:2019oqb}, and models coupling to the Standard Model with such low mass scales would already be observed. Even if a hidden sector model is capable of accommodating a very strong phase transition at a very low temperature, one expects a spectrum at PTA frequencies which has a low-frequency slope $\beta=3$~\cite{Caprini:2009fx} and hence $\gamma=2$, which is disfavoured by the data. While some exceptions from that scaling exist, they require either an extremely strong transition~\cite{Lewicki:2020jiv} or modification of cosmological expansion~\cite{Ellis:2020nnr}, both of which would be extremely difficult to realise at low temperatures without violating other bounds. 

{\bf Conclusions:}
We have analysed the GW spectra produced by cosmic string networks, recasting them numerically as power laws in the frequency range $f\in (2.5\times 10^{-9},1.2 \times10^{-8})$ Hz of interest to PTA experiments. This allowed us to express the resulting amplitude and slope as functions of the only free parameter in our model, which is the string tension $G\mu$. We then use these results to make contact with the recent NANOGrav 12.5~yr~\cite{Arzoumanian:2020vkk} data release, which finds evidence of {a stochastic common-spectrum
process, analysed in terms of power-law modelling,} that could be interpreted as a GW background.  We find that a cosmic string tension $G\mu\in (4\times 10^{-11},10^{-10})$ fits the data within the 68\% CL region around the best fit while $G\mu\in (2\times 10^{-11},3\times 10^{-10})$ is compatible with the data at the 95\% CL. Cosmic strings provide a better fit to the current data than a GW spectrum from SMBH mergers, which can fit the data at the 95\% CL but not the 68\% CL. We also show all next-generation GW detectors including SKA, LISA, TianQin, AEDGE, AION and ET will be able to probe the cosmic string spectra that fit the current data, whereas LIGO seems unlikely be able to probe them in the absence of additional cosmological or model features. 

A key probe of any GW interpretation of the NANOGrav data would be the appearance of quadrupole correlations, which have not (yet) been detected. Beyond this, measurement of a SGWB background compatible with the shape of spectrum shown in Fig.~\ref{fig:OmegaPTA} over a large range of frequencies would provide crucial confirmation of our bold GW interpretation of the NANOGrav 12.5~yr data.

{ \it Note added} - For completeness, we note that other models have also been 
proposed as possible explanations of the NANOGrav data, including 
\cite{Blasi:2020mfx, Buchmuller:2020lbh,Samanta:2020cdk, Chigusa:2020rks, Ramberg:2020oct} that also 
deal with cosmic strings, as well as explanations involving primordial 
black hole production \cite{DeLuca:2020agl,Vaskonen:2020lbd, Kohri:2020qqd, Sugiyama:2020roc, 
Domenech:2020ers,Bhattacharya:2020lhc}, cosmological phase transitions 
\cite{Nakai:2020oit,Neronov:2020qrl}, 
inflation \cite{Vagnozzi:2020gtf, Li:2020cjj, Kuroyanagi:2020sfw}, domain walls 
\cite{Bian:2020bps, Liu:2020mru, Chiang:2020aui}, audible axions \cite{Ratzinger:2020koh, Namba:2020kij}, and hypothetical violation of the null 
energy condition \cite{Tahara:2020fmn,Cai:2020qpu}.

%%%%%%%%%%%%%%%%%%%%%%%%%%%%%%%%%%%%%%%%%%%%%%%%%%%%%%%%%%%%%%%%%%%%
\acknowledgments 
This work was supported by the UK STFC Grant ST/P000258/1. J.E. also acknowledges support from the Estonian Research Council grant MOBTT5, and M.L. from the Polish National Science Center grant 2018/31/D/ST2/02048. The project is co-financed by the Polish National Agency for Academic Exchange within Polish Returns Programme under agreement PPN/PPO/2020/1/00013/U/00001

\bibliography{cosmicstrings}
\end{document}